\documentclass[aps,prx,superscriptaddress,twocolumn,floatfix,citeautoscript,longbibliography,hyperlinks]{revtex4-1}

\usepackage{graphicx}
\usepackage{dcolumn}
\usepackage{bm}
\usepackage[dvipsnames]{xcolor}
\usepackage[colorlinks=true,urlcolor=blue,citecolor=blue, linkcolor = blue]{hyperref}
\usepackage{amssymb,amsmath}
\newcommand{\bra}[1]{\left\langle #1\right|}
\newcommand{\ket}[1]{\left|#1\right\rangle}

\makeatletter
\newsavebox{\@brx}
\newcommand{\llangle}[1][]{\savebox{\@brx}{\(\m@th{#1\langle}\)}%
  \mathopen{\copy\@brx\kern-0.5\wd\@brx\usebox{\@brx}}}
\newcommand{\rrangle}[1][]{\savebox{\@brx}{\(\m@th{#1\rangle}\)}%
  \mathclose{\copy\@brx\kern-0.5\wd\@brx\usebox{\@brx}}}
\makeatother

\begin{document}

\title{Determination of dynamical quantum phase transitions in strongly correlated many-body systems using Loschmidt cumulants}

\author{Sebastiano Peotta}
\affiliation{Computational Physics Laboratory, Physics Unit, Faculty of Engineering and
Natural Sciences, Tampere University, FI-33014 Tampere, Finland}
\affiliation{Department of Applied Physics, Aalto University, FI-00076 Aalto, Finland}
\affiliation{Helsinki Institute of Physics, FI-00014 University of Helsinki, Finland}
\author{Fredrik Brange}
\affiliation{Department of Applied Physics, Aalto University, FI-00076 Aalto, Finland}
\author{Aydin Deger}
\affiliation{Department of Applied Physics, Aalto University, FI-00076 Aalto, Finland}
\author{Teemu Ojanen} \email{Email: teemu.ojanen@tuni.fi}
\affiliation{Computational Physics Laboratory, Physics Unit, Faculty of Engineering and
Natural Sciences, Tampere University, FI-33014 Tampere, Finland}
\affiliation{Helsinki Institute of Physics, FI-00014 University of Helsinki, Finland}
\author{Christian Flindt}
\affiliation{Department of Applied Physics, Aalto University, FI-00076 Aalto, Finland}

\begin{abstract}
Dynamical phase transitions extend the notion of criticality to non-stationary settings and are characterized by sudden changes in the macroscopic properties of time-evolving quantum systems. Investigations of dynamical phase transitions combine aspects of symmetry, topology, and non-equilibrium physics, however, progress has been hindered by the notorious difficulties of predicting the time evolution of large, interacting quantum systems. Here, we tackle this outstanding problem by determining the critical times of interacting many-body systems after a quench using Loschmidt cumulants. Specifically, we investigate dynamical topological phase transitions in the interacting Kitaev chain and in the spin-1 Heisenberg chain. To this end, we map out the thermodynamic lines of complex times, where the Loschmidt amplitude vanishes, and identify the intersections with the imaginary axis, which yield the real critical times after a quench. For the Kitaev chain, we can accurately predict how the critical behavior is affected by strong interactions, which gradually shift the time at which a dynamical phase transition occurs. We also discuss the experimental perspectives of predicting the first critical time of a quantum many-body system by measuring the energy fluctuations in the initial state, and we describe the prospects of implementing our method on a near-term quantum computer with a limited number of qubits. Our work demonstrates that Loschmidt cumulants are a powerful tool to unravel the far-from-equilibrium dynamics of strongly correlated many-body systems, and our approach can immediately be applied in higher dimensions.
\end{abstract}

\maketitle

\section{Introduction}

Whether or not quantum many-body systems out of equilibrium can be understood in terms of well-defined phases of matter is a central question in condensed matter physics. The lack of universal principles, such as those governing equilibrium systems \cite{Chandler:1987,Goldenfeld:2018}, makes the problem exceptionally hard. Still, the concepts of criticality and far-from-equilibrium dynamics have recently been elegantly unified through the discovery of dynamical phase transitions in which a time-evolving quantum many-body system displays sudden changes of its macroscopic properties~\cite{Heyl:2013,Karrasch2013,Andraschko2014,Vajna2014,Vajna2015,Budich2016,Zvyagin2016,Halimeh2017,Heyl:2018}. In equilibrium physics, phase transitions are reflected by singularities in the free energy, and dynamical phase transitions are similarly given by critical \emph{times}, where a non-equilibrium analogue of the free energy becomes non-analytic. Specifically, the role of the partition function is played by the return, or Loschmidt, amplitude of the many-body system after a quench, and its logarithm yields the corresponding free energy.

\begin{figure}[t]
    \centering
    \includegraphics[width=0.95\columnwidth]{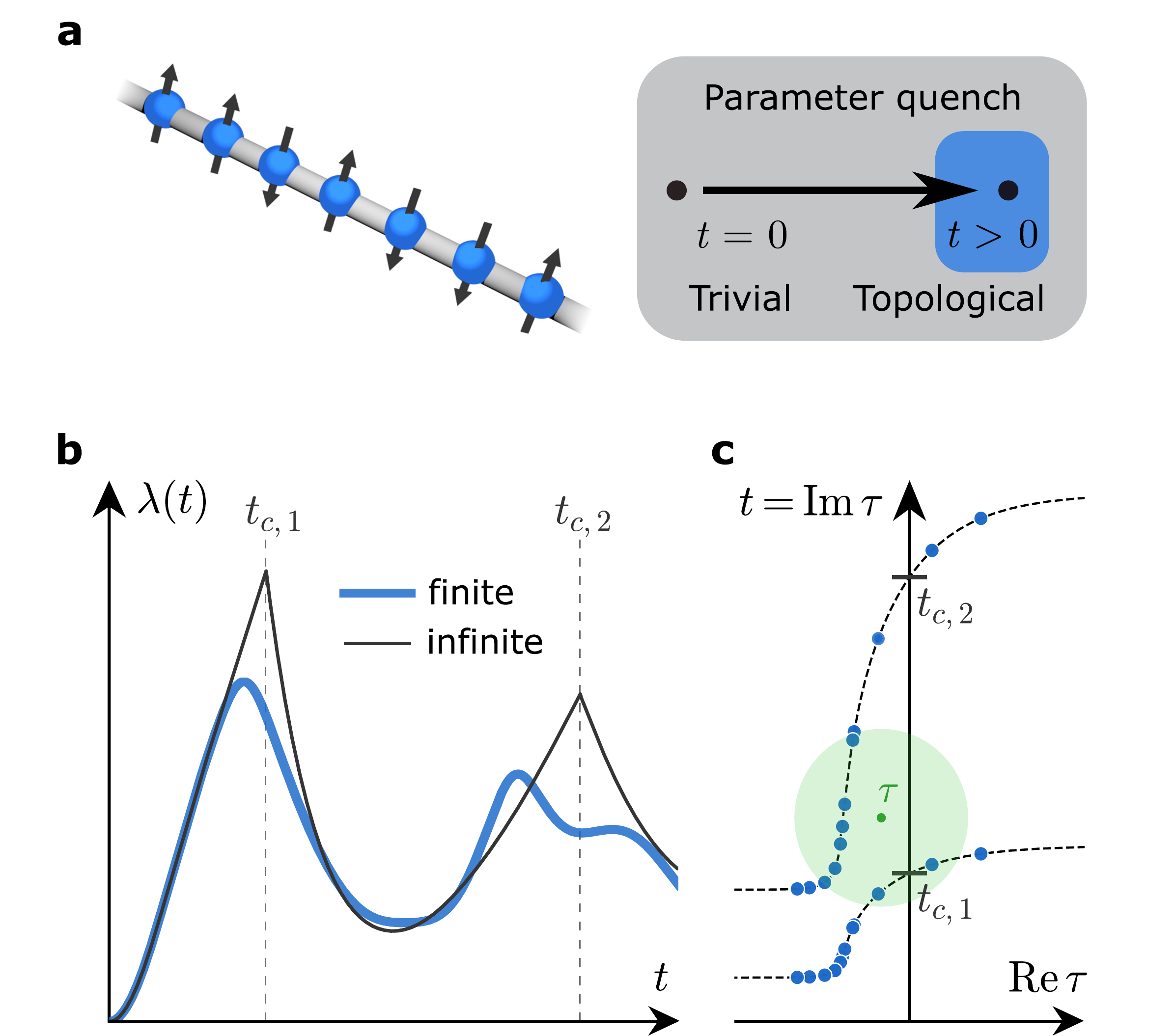}
    \caption{ \textbf{Dynamical phase transitions. a,} A sudden quench of the system parameters causes a dynamical phase transition in a quantum spin chain with $L$ sites. \textbf{b,} In the thermodynamic limit, such phase transitions give rise to singularities in the rate function at the critical times, $t_{c,1}, t_{c,2}\ldots $, see Eqs.~\eqref{eq:Loschmidt_amplitude} and \eqref{eq:ratefunc} for definitions, however, in finite-size systems they are smeared out. \textbf{c,} The singularities in the rate function are associated with the zeros (circles) of the Loschmidt amplitude in the complex-time plane. In the thermodynamic limit, they form continuous lines, and the real critical times are given by the crossing points with the imaginary axis. We determine the zeros of the Loschmidt amplitude from the Loschmidt cumulants evaluated at the basepoint $\tau$.}
    \label{fig:schematic}
\end{figure}

A typical setup for observing dynamical quantum phase transitions is depicted in Fig.~\ref{fig:schematic}\textbf{a}: a one-dimensional chain of interacting quantum spins is initialized in a ground state characterized by one type of order (or the lack of it) and subsequently made to evolve according to a Hamiltonian whose ground state possesses a different order. Experimentally, dynamical phase transitions have been observed in strings of trapped ions~\cite{Jurcevic2017,Zhang2017}, optical lattices~\cite{Flaeschner2018}, and several other systems that offer a high degree of control~\cite{Guo2019,Tian2019,Wang2019,Tian2020,Xu2020}. The Loschmidt amplitude is the overlap between the initial state of the system and the state of the system at a later time. Moreover, similarly to equilibrium systems, dynamical phase transitions may occur at critical times, where the Loschmidt amplitude vanishes, and the dynamical free energy becomes non-analytic in the thermodynamic limit. As illustrated in Fig.~\ref{fig:schematic}\textbf{b}, these non-analytic signatures may appear as cusps in the dynamical free energy, however, strictly speaking, they only occur for infinitely large systems. For finite-size systems, they are typically smeared out, and often for spin chains at least $L\simeq 50-100$ spins are required in order to identify and determine the critical times of a dynamical phase transition. Since the Hilbert space dimension grows exponentially with the chain length, the outstanding bottleneck for theoretical investigations of dynamical phase transitions is the need to predict the far-from-equilibrium dynamics of large quantum systems. Numerically, the task is computationally costly, or even intractable, and generally it requires advanced system-specific techniques that do not easily generalize to other systems or spatial dimensions~\cite{Pozsgay2013,Karrasch2013,Andraschko2014,Kriel2014,Sharma2015,Halimeh2017,ZaunerStauber2017,Homrighausen2017,Heyl:2018a,Kennes2018,Hagymasi:2019,Lacki2019,Huang2019,Halimeh2020,zunkovic2018,weidinger2017,Feldmeier2019}. For this reason, little is still known about dynamical phase transitions and the general applicability of concepts like universality and scaling. In fact, our current understanding comes to a large extent from a few exactly solvable models~\cite{Heyl:2013,Karrasch2013,Andraschko2014,Vajna2014,Heyl2015,Vajna2015,Schmitt2015,Budich2016,Zvyagin2016,Halimeh2017,Fogarty2017,bhattacharya2017,Heyl:2018,Kennes2018,Sedlmayr2018,Jafari2019,Defenu2019,Najafi2019,Gulacsi2020,zamani2020}. Important questions concern the relationship between critical times and dynamical changes in local observables or the entanglement spectrum (or other dynamical measures), which often exhibit similar but not strictly related time scales. However, case-by-case investigations have revealed that dynamical phase transitions are often accompanied by
interesting dynamics with comparable time scales, and one could view them as indicators of nontrivial dynamics in other many-body properties.

Here we pave the way for systematic investigations of dynamical phase transitions in correlated systems using Loschmidt cumulants, which allow us to accurately predict the critical times of a quantum many-body system using remarkably small system sizes, on the order of $L\simeq10-20$. Using modest computational power, we determine the critical times of the interacting Kitaev chain and the spin-1 Heisenberg chain after a quench and find for instance that a dynamical phase transition in the Kitaev chain gets suppressed with increasing interaction strength. The Loschmidt cumulants allow us to determine the complex zeros of the Loschmidt amplitude as illustrated in Fig.~\ref{fig:schematic}\textbf{c}. We can thereby map out the thermodynamic lines of zeros and identify the crossing points with the imaginary axis, corresponding to the real critical times, where a dynamical phase transition occurs. This approach makes it possible to predict the critical dynamics of a wide range of strongly interacting quantum many-body systems and is applicable also in higher dimensions. In two dimensions, the zeros can make up lines or surfaces in the complex plane, and our method can be used to determine all of these zeros as well as their density. Moreover, as we will show, our method provides exciting perspectives for future experiments on dynamical phase transitions. Specifically, our method makes it possible to predict the first critical time of a quantum many-body system after a quench by measuring the fluctuations of the energy in the initial state. In addition, due to the favorable scaling properties of our method, it is conceivable that it can be implemented on a near-term quantum computer with a limited number of qubits.

We now proceed as follows. In Sec.~\ref{sec:LC}, we develop our method for determining the zeros of the Loschmidt echo and their crossing points with the imaginary axis in the thermodynamic limit, which yield the critical times of a quantum many-body system after a quench. In Sec.~\ref{sec:Kitaev}, we consider dynamical phase transitions in the Kitaev chain after a quench, and we show how we can determine the critical times from remarkably small chain lengths even with strong interactions. Section~\ref{sec:heisenberg} is devoted to the spin-1 Heisenberg chain and includes several quenches, for instance, from the Haldane phase to the N\'eel phase. In Sec.~\ref{sec:exp}, we discuss the experimental perspectives for future realizations of our method. Finally, in Sec.~\ref{sec:concl}, we state our conclusions and provide an outlook on possible avenues for further developments. 

\section{From Loschmidt cumulants to Loschmidt zeros}
\label{sec:LC}

The fundamental object that describes dynamical phase transitions is the Loschmidt amplitude,
\begin{equation}
\label{eq:Loschmidt_amplitude}
\mathcal{Z}(it) = \bra{\Psi_0}e^{-it\mathcal{\hat H}}\ket{\Psi_0},
\end{equation}
where $\ket{\Psi_0}$ is the initial state of the many-body system at time $t=0$, the post-quench Hamiltonian $\mathcal{\hat H}$ governs the time evolution for times $t>0$, and we set $\hbar=1$ from now on. The Loschmidt amplitude resembles the partition function of a thermal system with Hamiltonian $\mathcal{\hat H}$, however, the inverse temperature is replaced by the imaginary time $\tau = it$, and an average is taken with respect to the initial state $\ket{\Psi_0}$. In equilibrium settings, a thermal phase transition occurs, if a system is cooled below its critical temperature, and it abruptly changes from a disordered to an ordered phase. Similarly, {\it dynamical} phase transitions occur at critical {\it times}, when a quenched system suddenly changes from one phase to another with fundamentally different properties. Such transitions are manifested in the rate function
\begin{equation}
\lambda(t)=-\frac{1}{L}\ln |\mathcal{Z}(it)|^2,
\label{eq:ratefunc}
\end{equation}
which is the non-equilibrium analogue of the free energy density. In some cases, dynamical phase transitions occur, if a system is quenched across an underlying equilibrium phase transition, however, generally, there is no simple relation between dynamical and equilibrium phase transitions. In Fig.~\ref{fig:schematic}\textbf{a}, the system size, denoted by $L$, is the total number of spins along the chain. In the thermodynamic limit of infinitely large systems, dynamical phase transitions give rise to singularities in the rate function, for example a cusp as shown in Fig.~\ref{fig:schematic}\textbf{b}. However, this non-analytic behavior typically becomes apparent for very large systems, and it is hard to pinpoint for smaller systems. For this reason, dynamical phase transitions are difficult to capture in computations and simulations, where the numerical costs grow rapidly with system size. 

Here we build on recent progress in Lee-Yang theories of thermal phase transitions~\cite{Deger:2018,Deger:2019,Deger:2020,Deger:2020b} and use Loschmidt cumulants to predict dynamical phase transitions in strongly-correlated many-body systems using remarkably small system sizes. The Lee-Yang formalism of classical equilibrium phase transitions considers the zeros of the partition function in the complex plane of the external control parameters~\cite{Yang:1952,Lee:1952,Blythe:2003,Bena:2005}. In a similar spirit, we treat the Loschmidt amplitude as a function of the complex-valued variable $\tau$. The Loschmidt amplitude of a finite system is an entire function, which can be factorized as~\cite{Arfken2012}
\begin{equation}
\label{eq:factorized_Z}
 \mathcal{Z}(\tau) = e^{\alpha\tau}\prod_{k}\left(1-\tau/\tau_k\right),
\end{equation}
where $\alpha$ is a constant, and $\tau_k$ are the complex zeros of the Loschmidt amplitude. For a thermal system, the values of the inverse temperature for which the partition function vanishes are known as Fisher zeros \cite{Fisher:1965}. We refer to the zeros of the Loschmidt amplitude as Loschmidt zeros. For a finite system, the zeros are isolated points in the complex plane. However, they grow denser as the system size is increased, and in the thermodynamic limit, they coalesce to form continuous lines and regions. Their intersections with the imaginary $\tau$ axis determine the real critical times, at which the rate function becomes non-analytic, and dynamical phase transitions occur~\cite{Heyl:2018}. As such, this phenomenology resembles the classical Lee-Yang theory of thermal phase transitions~\cite{Yang:1952,Lee:1952,Blythe:2003,Bena:2005}.

The central task is thus to determine the Loschmidt zeros. To this end, we define the Loschmidt moments and cumulants of the Hamiltonian $\mathcal{\hat H}$ of order $n$ as 
\begin{equation}
\label{eq:moment}
\langle \mathcal{\hat H}^n \rangle_\tau = (-1)^n\frac{\partial_{\tau}^n\mathcal{Z}(\tau)}{\mathcal{Z}(\tau)}
\end{equation}
and
\begin{equation}
\label{eq:cumulant}
\llangle \mathcal{\hat H}^n \rrangle_\tau = (-1)^n\partial_{\tau}^n\ln \mathcal{Z}(\tau),
\end{equation} 
where $\tau$ is the basepoint, at which the moments and cumulants are evaluated. For $\tau=0$, the Loschmidt moments reduce to the moments of the Hamiltonian with respect to the initial state as $\langle \mathcal{\hat H}^n \rangle_0 = \bra{\Psi_0}\mathcal{\hat H}^n \ket{\Psi_0}$. At finite times, the Loschmidt moments are $\langle \mathcal{\hat H}^n \rangle_\tau = \bra{\Psi_0}\mathcal{\hat H}^n \ket{\Psi(\tau)}/\langle\Psi_0 |\Psi(\tau)\rangle$, where $\ket{\Psi(\tau)} =e^{-\tau\mathcal{\hat H}}\ket{\Psi_0}$ is the time-evolved state. The cumulants can be obtained from the moments using the standard recursive formula
\begin{equation}
\llangle \mathcal{\hat H}^n \rrangle_\tau = \langle \mathcal{\hat H}^n \rangle_\tau - \sum_{m = 1}^{n-1}\binom{n-1}{m-1}\llangle \mathcal{\hat H}^m \rrangle_\tau \langle \mathcal{\hat H}^{n-m} \rangle_\tau.
\end{equation}

For our purposes, it is now convenient to define the normalized Loschmidt cumulants $\kappa_n(\tau)$ as
\begin{equation}
\label{eq:cumulant_series}
\kappa_n(\tau) = \frac{(-1)^{n-1}}{(n-1)!}\llangle \mathcal{\hat H}^n \rrangle_\tau = \sum_{k}\frac{1}{(\tau_k-\tau)^n},\quad  n>1,
\end{equation}
having used Eq.~\eqref{eq:factorized_Z} to express them in terms of the zeros. This expression shows that the Loschmidt cumulants are dominated by the zeros that are closest to the (complex) basepoint $\tau$, while the contributions from other zeros rapidly fall off with their inverse distance from the basepoint to the power of the cumulant order $n$. The main idea is now to extract the $m$ closest zeros from $2m$ high Loschmidt cumulants, which we can calculate. For $m=2$, this can be done by adapting the method from Refs.~\cite{Deger:2018,Deger:2019,Deger:2020,Deger:2020b}. However, for arbitrary $m$, we use the general approach presented in App.~\ref{app:determination_zeros}-\ref{app:errors}. For the systems that we consider in the following, we extract the $m = 7$ zeros closest to the movable basepoint using Loschmidt cumulants of order $n=9$ to $n=22$. 

It should be emphasized that our approach hardly makes any assumptions about the quantum many-body system at hand or the method used for obtaining the cumulants. As outlined in App.~\ref{app:Krylov_method}, we use a Krylov subspace method~\cite{Park1986,Paeckel2019} to perform the complex time evolution and evaluate the Loschmidt moments and cumulants, which we then use for extracting the Loschmidt zeros. All results presented below have been obtained on a standard laptop, and the method can readily be adapted to a variety of interacting quantum many-body systems, also in higher dimensions.

\begin{figure*}[ht]
    \centering
    \includegraphics[]{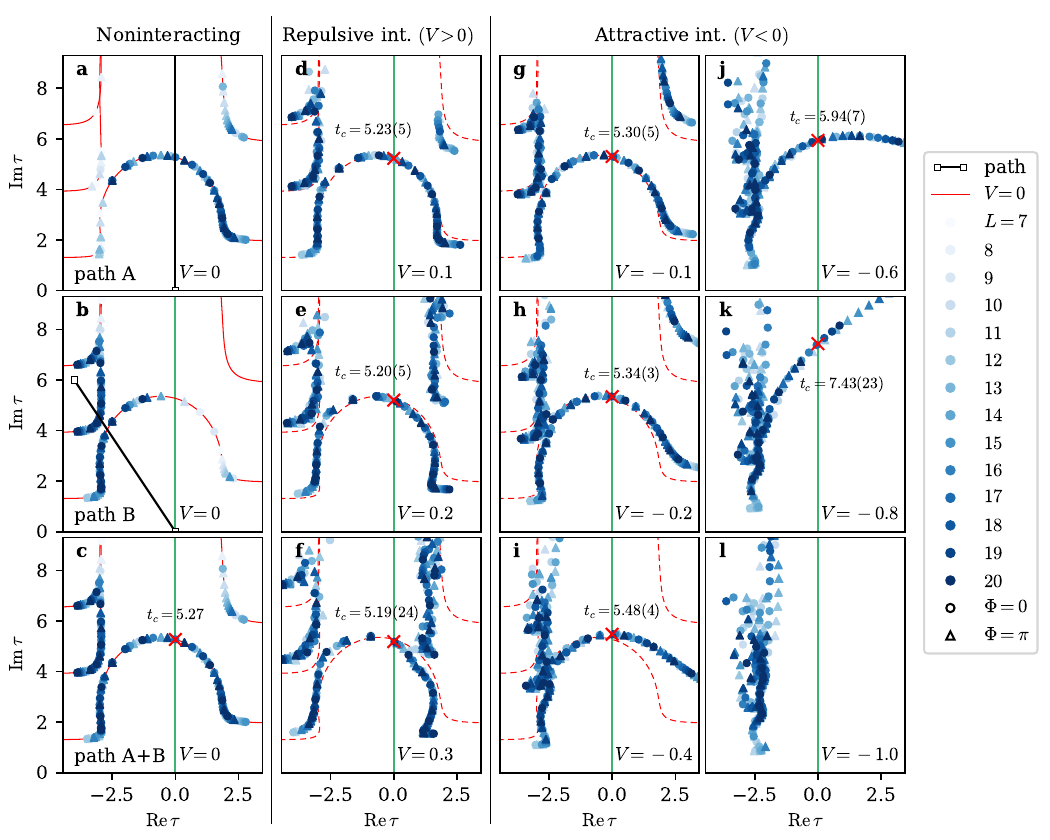}
    \caption{{\bf The interacting Kitaev chain.} We quench the chemical potential from the trivial phase $\mu=-1.4$ to the topological phase $\mu=-0.2$ for $t>0$ with fixed $\Delta=0.3$ (all parameters and the inverse time $\tau^{-1}$ are expressed in units of $J=1$). \textbf{a-c,} Complex zeros for different system sizes ($L=7-20$) and boundary conditions ($\Phi = 0,\pi$) in the noninteracting case, compared with the exact thermodynamic lines of zeros. Only the zeros within a finite range from the basepoint can be accurately obtained from the Loschmidt cumulants. This fact is illustrated by moving the basepoint $\tau$ along two different paths (paths A and B in panels \textbf{a} and \textbf{b}). Panel \textbf{c} combines the results from panels \textbf{a} and \textbf{b}. \textbf{d-f,} Loschmidt zeros and critical times obtained with repulsive interactions ($V >0$) along the two paths. The lines of zeros for $V = 0$ (dashed line) are shown for comparison. The critical time $t_c$, shown in each panel as a red cross, is obtained as the intersection between the imaginary axis and the line drawn from the zero $\tau_-$ with smallest negative real part  (in absolute value) to the zero $\tau_+$ with the smallest positive real part. The error on the critical time is estimated as $\Delta t_c = \max(|t_c-\mathrm{Im}\,\tau_-|,|t_c-\mathrm{Im}\,\tau_+|)$. \textbf{g-l,} Evolution of zeros and critical times with increasing attractive interactions ($V <0$). For very strong interactions ($V = -1$, panel \textbf{l}), the zeros do not cross the imaginary axis, signalling the absence of a dynamical quantum phase transition. As discussed in App.~\ref{app:errors}, the numerical error in the zeros is of the order of $10^{-3}$.}
    \label{fig:kitaev}
\end{figure*}

\section{Interacting Kitaev chain}
\label{sec:Kitaev}
 
We first consider the spin-1/2 XYZ chain or, equivalently, the interacting Kitaev chain. The non-interacting limit maps to the XY model, which was solved exactly in the pioneering work of Ref.~\cite{Heyl:2013}. Here, we use Loschmidt cumulants to predict a dynamical quantum phase transition in the {\it strongly interacting} regime. The Hamiltonian of the spin-1/2 XYZ chain with a Zeeman field reads
\begin{equation} \label{eq:spinhalf}
\mathcal{\hat H} =\sum_{\alpha,j = 1}^{L} J_\alpha\hat S_j^\alpha\hat S_{j+1}^\alpha
- h\sum_{j = 1}^L\hat{S_j^z}\,,
\end{equation} 
where $\hat{S}_j^\alpha$ are the spin-1/2 operators for the $\alpha=x,y,z$ component of the spin on site $j$ of the chain of length $L$, the exchange couplings are denoted by $J_\alpha$, and $h$ is the Zeeman field. We use twisted boundary conditions, 
\begin{equation}
\label{eq:bound_cond}
\begin{split}
\hat{S}_{L+1}^x&=\cos(\Phi)\hat{S}_1^x+\sin(\Phi)\hat{S}^y_1,\\
\hat{S}_{L+1}^y&=-\sin(\Phi)\hat{S}_1^x+\cos(\Phi)\hat{S}^y_1,
\end{split}
\end{equation}
and $\hat{S}_{L+1}^z =\hat{S}_{1}^z$, where $\Phi$ is the twist angle. In the fermionic representation, obtained by a Jordan-Wigner transformation \cite{Franchini_book}, the model maps to the interacting Kitaev chain of spinless fermions with operators $\hat{c}_j$~and~$\hat{c}_{j}^\dagger$,
\begin{widetext}
\begin{equation}
\label{eq:kitaev}
\mathcal{\hat{H}} = -\frac{1}{2}\sum_{j = 1}^{L-1}\left(J\,\hat {c}_{j}^\dagger\hat{c}_{j+1} + \Delta\, \hat {c}_{j}^\dagger\hat{c}_{j+1}^\dagger + \text{H.c.}\right)
+V\sum_{j = 1}^{L-1} \left(\hat {n}_{j}-\frac{1}{2}\right)\left(\hat {n}_{j+1}-\frac{1}{2}\right)-\mu\sum_{j = 1}^{L} \hat {c}_{j}^\dagger\hat{c}_{j}
+\frac{s_\Phi \hat{P}}{2} \left(J\,\hat {c}_{L}^\dagger\hat{c}_{1} + \Delta\, \hat {c}_{L}^\dagger\hat{c}_{1}^\dagger + \text{H.c.}\right),
\end{equation}
\end{widetext}
where the twist angle now enters in the last line through the parameter $s_\Phi$, which is $+1$, if the twist angle is $\Phi = 0$, and $-1$, if $\Phi = \pi$. These are the only two values of the twist angle used here. The parameters of the two Hamiltonians are related as $J=-(J_x+J_y)/2$, $\Delta=(J_y-J_x)/2$, $\mu=-h$, and $V=J_z$. Moreover, the number operator on site $j$ is $\hat{n}_j = \hat{c}_j^\dagger\hat{c}_j$, while $\hat{P} = \exp(i\pi\sum_{j}\hat{n}_j)$ is the parity operator. 

The Kitaev chain describes a one-dimensional superconductor with a $p$-wave pairing term that is proportional to $\Delta$, supporting two distinct topological phases. The two values of the twist angle, $\Phi = 0, \pi$, physically correspond to a magnetic flux equal to zero or half flux quantum threaded through the ring-shaped chain. These are the only distinct flux values that are consistent with superconducting flux quantization. It is useful to vary the boundary conditions, since in the non-interacting case ($V = 0$), which corresponds to the exactly solvable spin-1/2 XY model, the Loschmidt zeros can be labeled by the quasi-momentum $k_m = (2\pi m - \Phi)/L$ with $m = 0, \dots,L-1$ \cite{Heyl:2013}. Thus, by using the two different values of $\Phi$, we can sample the thermodynamic line of zeros twice as densely for a given system size. It turns out that even in the interacting case ($V\neq 0$) it is useful to vary the boundary conditions for the same reason.
 
We are now ready to investigate dynamical quantum phase transitions in the interacting Kitaev chain. To this end, we take for the initial state $\ket{\Psi_0}$ the ground state of the Hamiltonian~\eqref{eq:kitaev} with $|\mu /J| > 1$, which corresponds to the topologically trivial phase, and perform a quench into the topological regime with $|\mu/J|<1$ for later times, $t>0$. As shown in Fig.~\ref{fig:kitaev}, from the Loschmidt cumulants we can find the complex zeros of the Loschmidt amplitude even with attractive ($V < 0$) or repulsive ($V >0$) interactions, for which an analytic solution is not available. In the left column, we first consider the non-interacting case, where the thermodynamic lines of zero can be determined analytically~\cite{Heyl:2013}. In panels \textbf{a} and \textbf{b}, we show the zeros found from the Loschmidt cumulants as the basepoint $\tau$ is moved along the paths denoted by A and B, respectively, while panel \textbf{c} shows the combined results. Remarkably, the Loschmidt cumulants allow us to map out the thermodynamic lines of zeros using chains of rather short lengths, $L=7-20$, and thereby identify the crossing points with the imaginary axis, corresponding to the real critical times, where a dynamical phase transition occurs. The comparison between the exact and the approximate zeros obtained from the Loschmidt cumulants provides an important estimate of the accuracy. In the worst cases, the accuracy is an order of magnitude better than the size of the markers in Fig.~\ref{fig:kitaev} (see App.~\ref{app:errors}). We note that our choice of the paths in Fig.~\ref{fig:kitaev} was guided by our knowledge of the zeros in the non-interacting case. However, more generally, without any specific knowledge of a system, one may choose paths that scan the complex plane, in particular along the imaginary
time axis and its immediate vicinity (since those zeros determine whether and when the system exhibits a dynamical phase transition).

Having benchmarked our approach in the non-interacting case, we move on to the strongly interacting regime. In the second column of Fig.~\ref{fig:kitaev}, we show the Loschmidt zeros for repulsive interactions, which tend to shift the critical crossing point with the imaginary axis to earlier times. A more dramatic effect is observed in the third and fourth columns, where we gradually increase the attractive interactions. In this case, the dynamical phase transition happens at later times, and eventually, for sufficiently strong interactions, the thermodynamic line of zero no longer crosses the imaginary axis, implying the absence of a dynamical phase transition. 

While in the noninteracting limit the small systems reproduce the thermodynamic lines essentially exactly, interactions give rise to finite-size effects when two different lines come close. Despite that, sufficiently isolated lines and segments, such as the ones that determine the dynamical phase transitions in Fig.~\ref{fig:kitaev}, remain scale-invariant. We stress that these results are obtained for very small chains of lengths from $L=10$ to $L=20$, which, while remarkable, is in line with similar observations for the Lee-Yang zeros in classical equilibrium systems \cite{Deger:2018,Deger:2019,Deger:2020,Deger:2020b}. In particular, for strongly interacting systems, the use of such system sizes makes the approach very attractive from a computational point of view, since direct calculations of the Loschmidt amplitude typically require system sizes that are an order of magnitude larger, in generic cases with an exponential increase in the computational cost.

\begin{figure*}
    \centering
    \includegraphics[]{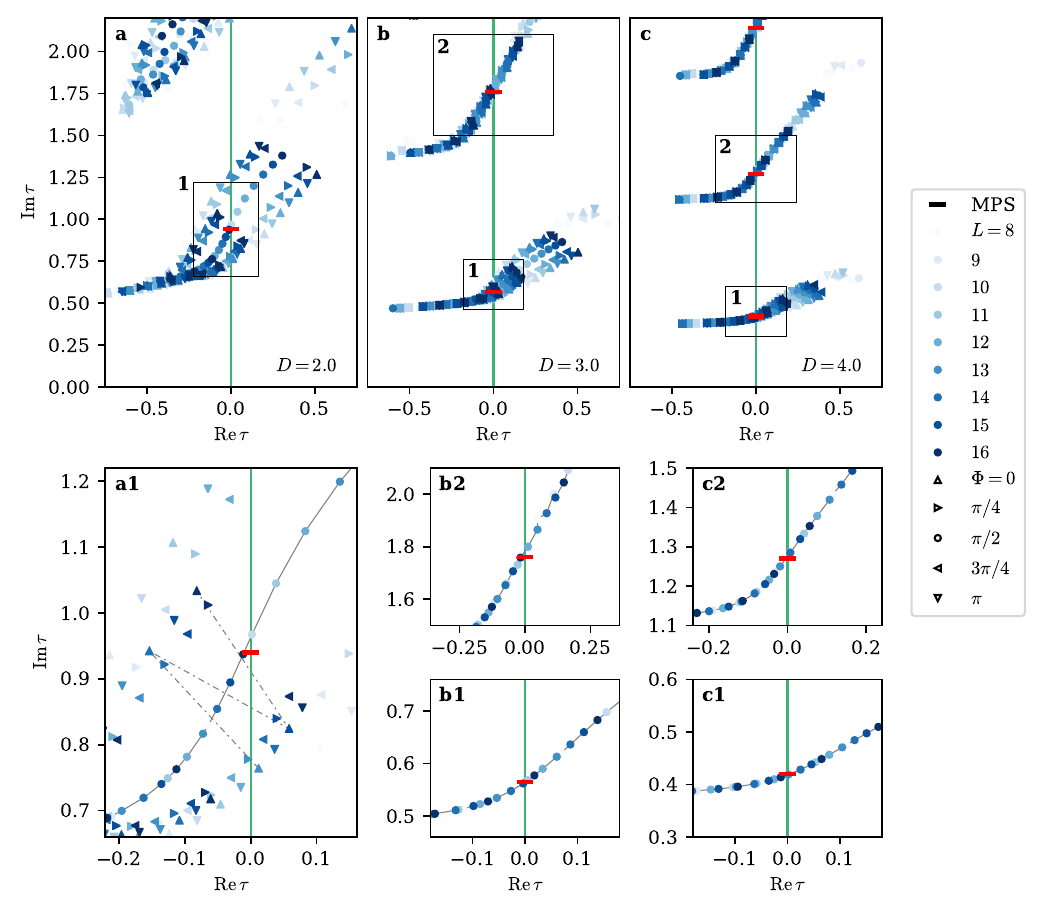}
    \caption{{\bf The Heisenberg chain.} We quench the system from the Haldane phase to the large-$D$ phase. The initial state is the ground state of the model \eqref{eq:spinone} with $J=J_z=3B$ and $D=0$ (the AKLT state \cite{Affleck1987}). For the post-quench Hamiltonian, we set $B = 0$, while $D$ can take the values 2, 3, or 4. Here $J=J_z=1$ is the unit of energy and inverse time. \textbf{a}, Loschmidt zeros for $D=2$. Panel \textbf{a1} is a magnified view of the area within the black rectangle in panel \textbf{a}. From panels \textbf{a} and \textbf{a1} one can clearly see how the position of a Loschmidt zero for fixed $L$ depends on the twist angle $\Phi$, which is a finite-size effect. It is also useful to consider a fixed twist angle and vary the system size as in the case of the zeros connected by the dash-dotted line in panel \textbf{a1} ($\Phi = 0$, $L = 13$, $14$, $15$, $16$). The finite-size dependence is suppressed for the zeros corresponding to the twist angle $\Phi=\pi/2$, defining the effective thermodynamic line of zeros (solid line, see App.~\ref{app:boundary_conditions}). The critical time, determined by the crossing of the effective line with the imaginary axis, is in excellent agreement with the result of Ref.~\cite{Hagymasi:2019} (red bar) obtained using matrix product states (MPS). \textbf{b}, \textbf{c} Same as in panel \textbf{a} but with $D=3,4$. Finite-size effects are suppressed with increasing $D$. In panels \textbf{b1}, \textbf{b2}, \textbf{c1}, \textbf{c2} the crossings of the effective thermodynamic lines of zeros with the imaginary axis are shown and compared again with the critical times obtained in Ref.~\cite{Hagymasi:2019}.}
    \label{fig:spin1_1}
\end{figure*}

\section{Spin-1 Heisenberg chain}
\label{sec:heisenberg}

The Kitaev chain from above possesses an exactly solvable limit, which provides an important benchmark for the use of the Loschmidt cumulants. However, generically, exact solutions are not available, which makes the usefulness of the Loschmidt cumulants further evident. For this reason, we now consider the spin-1 Heisenberg chain, which harbors a rich phase diagrams both with symmetry-broken phases and a topological phase, the Haldane phase \cite{Chen:2003}. The spin-1 Heisenberg chain is defined by the Hamiltonian 
\begin{equation} \label{eq:spinone}
\begin{split}
\mathcal{\hat H} = &\sum_{j = 1}^{L}\left[J\big(\hat S_j^x\hat S_{j+1}^x+\hat S_j^y\hat S_{j+1}^y\big) +J_z\hat S_j^z\hat S_{j+1}^z\right]\\ &+ D\sum_{j = 1}^L(\hat{S_j^z})^2+B\sum_{j = 1}^{L}(\hat{ \mathbf{S}}_j\cdot\hat{ \mathbf{S}}_{j+1})^2\,,
\end{split}
\end{equation}
where $\hat S^{i}$ are spin-1 operators, the exchange couplings between neighboring spins are denoted by $J$ and $J_z$, while $D$ and $B$ characterize the single-spin uniaxial anisotropy and the biquadratic exchange coupling, respectively. The first line defines the spin-1 XXZ model, while for $J = J_z = 3B$ and $D = 0$, one obtains the Affleck-Kennedy-Lieb-Tasaki (AKLT) model, whose ground state is known explicitly \cite{Affleck1987}, despite the fact that the Hamiltonian~\eqref{eq:spinone} is not exactly solvable. Again, we employ twisted boundary conditions as defined in Eq.~\eqref{eq:bound_cond} for five different values of the phase $\Phi = 0,\, \pi/4,\, \pi/2,\, 3\pi/4,\, \pi$.
We consider two kinds of quenches in which different parameters in the Hamiltonian~\eqref{eq:spinone} are abruptly changed at $t=0$.

\begin{figure*}
    \centering
    \includegraphics[width=1.99\columnwidth]{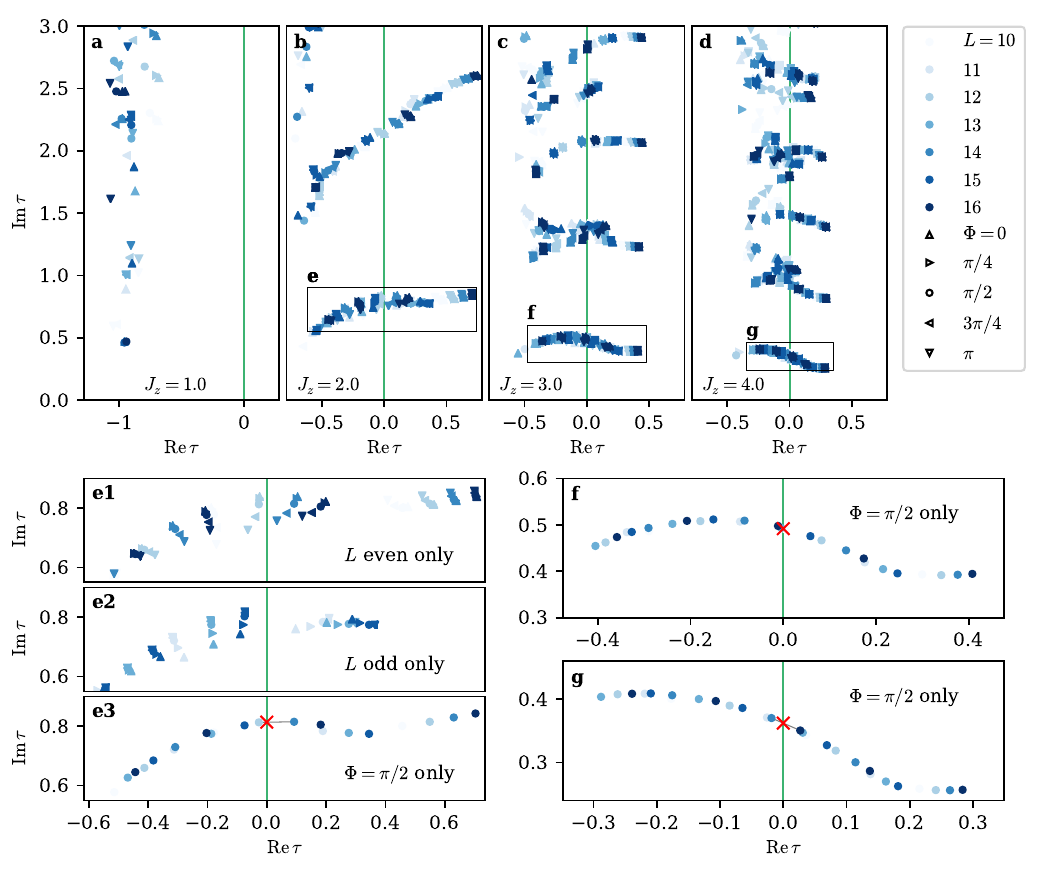}
    \caption{\textbf{Quench from the Haldane phase to the N\'eel phase.} The initial state is the ground state of the model \eqref{eq:spinone} with $J_z/J=1/2$ and $D=B=0$. The quench is performed by changing the parameter $J_{z}$ to the values $J_{z} = 1,\,2,\,3,\,4$ (in units of $J$). \textbf{a,} Loschmidt zeros for the quench to $J_{z}=1$, which is not sufficiently large to reach the N\'eel phase. In this case, the zeros do not cross the imaginary axis, and no dynamical phase transition occurs. \textbf{b}, \textbf{c}, \textbf{d} Similar to panel \textbf{a}, but with $J_z=2,3,4$. In this case, several dynamical phase transitions occur as shown for example in panels \textbf{e1-3} and \textbf{f}-\textbf{g}. The critical times are shown as red crosses and are estimated as done in Fig.~\ref{fig:spin1_1} using the zeros for $\Phi = \pi/2$ only (see App.~\ref{app:boundary_conditions} for details).}
    \label{fig:spin1_2}
\end{figure*}

In the first quench, we initialize the system in the AKLT ground state, which is a representative of the topological Haldane phase, and evolve it with the Hamiltonian~\eqref{eq:spinone} using the parameters $B = 0$, $J = J_z >0$ and $D/J = 2,\,3,\,4$. The ground states of the post-quench Hamiltonians are within the topologically trivial large-$D$ phase. The same quenches have been explored in Ref.~\cite{Hagymasi:2019} for system sizes up to $L=120$ using matrix product states, providing us with an important benchmark.

Figure~\ref{fig:spin1_1} shows the Loschmidt zeros for finite system sizes extracted from the Loschmidt cumulants for the first quench. We use twisted boundary conditions to gauge finite-size effects as the position of the Loschmidt zeros is expected to become insensitive to the phase $\Phi$ for very large systems. By contrast, for the relatively small system sizes used in Fig.~\ref{fig:spin1_1}, finite-size effects are pronounced in particular in panel \textbf{a}, which shows the zeros for the $D/J = 2$ quench. This value is the closest to the critical one $D_c/J \simeq 1$ (with $J = J_z$ and $B = 0$) separating the Haldane phase from the large-$D$ phase \cite{Chen:2003}, providing a plausible reason for the enhanced finite-size effects. Importantly, as discussed in App.~\ref{app:boundary_conditions}, the oscillatory pattern of zeros for different system sizes and twist angles is highly regular, which enables us to filter out the finite-size effects. In this prescription, a thermodynamic line of zeros is approximated by the smooth line of zeros emerging at the twist angle $\Phi=\pi/2$. 

The critical times of the transition, obtained from the crossings of the thermodynamic lines of zeros with the imaginary axis (see panels \textbf{a1}, \textbf{b1-2} and \textbf{c1-2} in Fig.~\ref{fig:spin1_1}), are in excellent agreement with the critical times obtained directly from the Loschmidt amplitude that was calculated using state-of-the-art computations in Ref.~\cite{Hagymasi:2019}. However, in contrast to Ref.~\cite{Hagymasi:2019}, which considers nearly an order of magnitude larger systems, our results are obtained from chain lengths up to $L=16$. This comparison provides an illustration of the power of our method in treating strongly correlated many-body systems.

While the exact correspondence between dynamical phase transitions and the equilibrium phase transitions of the respective model remains unknown \cite{Heyl:2018}, dynamical phase transitions are often observed, when the ground states of the initial and final Hamiltonians belong to different equilibrium phases. To explore this general scenario in the case of transitions between a topological phase and a symmetry-broken phase in a strongly correlated system, we solve for the first time quenches between the topological Haldane phase and the symmetry-broken N\'eel phase~\cite{Chen:2003}. In Fig.~\ref{fig:spin1_2}, we depict the Loschmidt zeros for the initial state with $D=B=0$ and quenching $J_z$ from $J_{z}/J = 1/2$ to the final values $J_{z}/J = 1,\,2,\,3,\,4$. The equilibrium quantum phase transition occurs at the critical value $J_{z,c} \simeq 1.2J$~\cite{Chen:2003}. Indeed, our results confirm that no dynamical phase transition is observed when $J_z/J = 1$, since all the Loschmidt zeros have negative real part as shown in panel \textbf{a} of Fig.~\ref{fig:spin1_2}. By contrast, for the other final values of $J_{z}$, which would put the equilibrium system in the antiferromagnetic N\'eel phase, dynamical phase transitions are observed. As in the first quench, finite-size effects are suppressed for quenches, where the final state resides deeper in the gapped phase. 

As we see in panels \textbf{e1-3} of Fig.~\ref{fig:spin1_2}, the Loschmidt zeros for different system sizes and boundary conditions have a structure similar to the one observed in the Haldane-to-large-$D$ quench. The same prescription as above irons out the finite-size oscillations and results in a smooth approximation of the thermodynamic lines of zeros. The critical times can then be accurately read off from the data obtained for chain lengths of $L \leq 16$, as in the case of the first quench.

\section{Experimental perspectives}
\label{sec:exp}
In the previous sections, we focused on using the Loschmidt cumulants for predicting dynamical phase transitions based on numerical calculations. However, as we will now discuss, our approach also provides perspectives for future experiments. We will show that it is possible to predict the first critical time of a quantum many-body system by measuring the fluctuations of the energy in the initial state. We will also discuss the prospects of implementing our method on a near-term quantum computer with a small number of qubits.

\begin{figure}
    \centering
    \includegraphics[width=0.96\columnwidth]{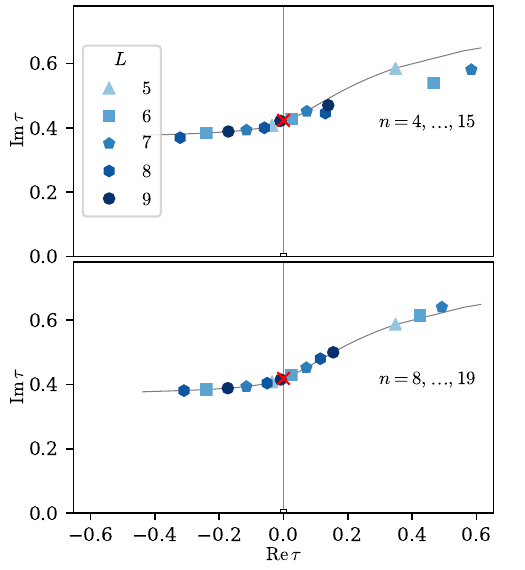}
    \caption{\textbf{Determination of the critical time from the initial energy fluctuations.} Loschmidt zeros for the Heisenberg chain~\eqref{eq:spinone} obtained from the energy fluctuations in the ground state of the model for $J=J_z=3B$ and $D=0$ at the initial time $\tau=0$, while the energy is determined by the post-quench Hamiltonian with $B = 0$ and $D=4$. Here $J=J_z=1$ is the unit of energy and inverse time. The zeros correspond to chains of lengths $L=5,\ldots,9$, and in the upper (lower) panel we have extracted the zeros using energy cumulants of orders $n=4,\ldots,14$ ($n=8,\ldots,19$). Importantly, the zeros converge to their exact positions with increasing cumulant orders as it can be seen by comparing the panels. The grey line corresponds to the zeros in  panel \textbf{c1} of Fig.~\ref{fig:spin1_1}, and the estimate of the critical time is indicated with a red cross. }
    \label{fig:exp}
\end{figure}

The Loschmidt moments are generally complex-valued, and it is not obvious how they can be measured. However, at the initial time, $\tau=0$, the Loschmidt moments simplify to the moments of the post-quench Hamiltonian with respect to the initial state as $\langle \mathcal{\hat H}^n \rangle_0 = \bra{\Psi_0}\mathcal{\hat H}^n \ket{\Psi_0}$. Thus, by repeatedly preparing the system in the state $\ket{\Psi_0}$ and measuring the energy given by the post-quench Hamiltonian $\mathcal{\hat H}$, one can construct the distribution of the energy and extract the corresponding moments and cumulants. From the cumulants, it is then possible to extract the closest Loschmidt zeros as demonstrated in Fig.~\ref{fig:exp} following a quench in a Heisenberg chain of lengths $L=5,\ldots,9$. From these results, we predict the critical time to be around $t_c\simeq 0.42$ as indicated by a red cross. This perspective is fascinating: by measuring the initial energy fluctuations, it is possible to predict the \emph{later} time at which a dynamical phase transition will occur. 

The idea behind such an experiment does not depend in detail on the actual physical implementation, and from a practical point of view different platforms may provide certain advantages. We expect, for example, that an experiment could be realized with atoms in optical lattices~\cite{Eckardt_2017} or with spin chains on surfaces \cite{Choi_2019}, systems that both offer a high degree of control and flexibility. As illustrated in Fig.~\ref{fig:exp}, it will be necessary to measure the high cumulants of the energy fluctuations. For large systems, accurate measurements of high cumulants are challenging, since the central-limit theorem dictates that distributions tend to be Gaussian with nearly vanishing high cumulants. However, for the small systems that we consider, the situation is different, and several quantum transport experiments have measured cumulants of up to order 20 \cite{Flindt_2009,Fricke_2010} and used them  for determining the zeros of generating functions \cite{Flindt_2013,Brandner_2017}, which are similar to the Loschmidt amplitude. Thus, an experimental determination of Loschmidt zeros for small interacting quantum systems appears feasible with current technology.

Our method may also be implemented on small near-term quantum computers, which are now becoming available. Such quantum computers allow for the specific tailoring of any desired Hamiltonian and for time-evolving an initial state both in real and imaginary time~\cite{Sun_2021,Lin_2021}. Thus, it will be possible to evaluate a time-evolved state of the form  $\ket{\Psi(\tau)} =e^{-\tau\mathcal{\hat H}}\ket{\Psi_0}$ and subsequently calculate the Loschmidt moments $\langle \mathcal{\hat H}^n \rangle_\tau = \bra{\Psi_0}\mathcal{\hat H}^n \ket{\Psi(\tau)}/\langle\Psi_0 |\Psi(\tau)\rangle$ and the corresponding cumulants from which the Loschmidt zeros are obtained. Again, the favorable scaling properties of our method become important, as they make it possible to predict the critical times of a quantum many-body system with only 10 to 20 constituents. Such sizes can soon be simulated on quantum computers with a limited number of qubits. 

\section{Conclusions}
\label{sec:concl}
 
We have demonstrated that Loschmidt cumulants are a powerful tool to unravel dynamical phase transitions in strongly interacting quantum many-body systems after a quench, making it possible to accurately predict the critical times of a quantum many-body system using remarkably small system sizes. Using modest computational power, we have explored dynamical phase transitions in the Kitaev chain and the spin-1 Heisenberg chain with a specific focus on the role of strong interactions. As we have shown, our approach circumvents the existing bottleneck of computing the full non-equilibrium dynamics of large quantum many-body systems, and instead we track the zeros of the Loschmidt amplitude in the complex plane of time in a similar spirit to the classical Lee-Yang theory of equilibrium phase transitions. As such, our approach paves the way for systematic investigations of the far-from-equilibrium properties of interacting quantum many-body systems, and we foresee many exciting perspectives ahead. In particular, our method can immediately be applied to dynamical phase transitions in dimensions higher than one, and the ease of implementing it may be critical for comprehensive investigations of the finite-size scaling close to a dynamical phase transition. We have also shown that our approach paves the way for exciting experimental developments by making it possible to predict the first critical time of a quantum many-body system in the thermodynamic limit by measuring the initial energy fluctuations in a much smaller system. In addition, due to the favorable scaling of our method, it seems feasible that it can be implemented on a near-term quantum computer with a limited number of qubits. In a broader perspective, the advances presented here may not only be useful for understanding the dynamical non-equilibrium properties of large quantum systems. They may also be helpful in designing novel quantum materials with specific, desired properties.

\acknowledgements
We thank the authors of Ref.~\cite{Hagymasi:2019} for providing us with their results for the spin-1 Heisenberg chain, which we used to extract the critical times indicated in Fig.~\ref{fig:spin1_1}. The work was supported by the Academy of Finland through the Finnish Centre of Excellence in Quantum Technology (project numbers 312057 and 312299) as well as projects number 308515, 330384, 331094, and 331737. F.B.~acknowledges support from the European Union's Horizon 2020 research and innovation programme under the Marie Sk\l{}odowska-Curie grant agreement number 892956. T.O.~acknowledges project funding from Helsinki Institute of Physics. 

\appendix

\section{Determination of Loschmidt zeros}
\label{app:determination_zeros}

Here we present the basic idea of the method used to extract the Loschmidt zeros from the Loschmidt cumulants. More details on the specific procedure employed to obtain the results shown in Figs.~\ref{fig:kitaev}-\ref{fig:spin1_2} are provided in App.~\ref{app:errors}, where we for instance explain how we estimate the error that affects the approximate zeros extracted with our method. 

In Eqs.~(\ref{eq:factorized_Z},\ref{eq:cumulant_series}), repeated zeros are allowed such that every distinct zero appears in the series~\eqref{eq:cumulant_series} as many times as its multiplicity. It is convenient in the following to use a different labelling of the Loschmidt amplitude zeros in which the index $k$ runs over the distinct zeros ($\tau_k \neq \tau_{k'}$ for $k\neq k'$), and also to denote by $d_k$ the multiplicity of the $k$-th zero. Using this convention, Eq.~\eqref{eq:cumulant_series} becomes
\begin{equation}
\label{eq:kappa_n_trunc}
\kappa_n(\tau)
=\sum_{k=0}^{\infty}\frac{d_k}{(\tau_k-\tau)^n} \simeq \sum^{m-1}_{k = 0} d_k\lambda_k^n\,.
\end{equation} 
On the right hand side, we have introduced $\lambda_k = 1/(\tau_k-\tau)$ and truncated the sum to the $m$ zeros closest to the basepoint. This is a good approximation for large $n$ since the contribution of each zero to the normalized cumulant $\kappa_n(\tau)$ is suppressed by its inverse distance to the basepoint raised to the power of the cumulant order.

We now determine the zeros of the Loschmidt amplitude based on the fact that, if Eq.~\eqref{eq:kappa_n_trunc} were exact, the normalized cumulants would satisfy a homogeneous linear difference equation of degree $m$ of the form
\begin{equation} \label{eq:diff_eq}
\kappa_n = a_1\kappa_{n-1}+a_2\kappa_{n-2}+\ldots + a_m 
\end{equation}
for some coefficients $a_l$.
Indeed, the general solution of Eq.~\eqref{eq:diff_eq} is given by the right hand side of Eq.~\eqref{eq:kappa_n_trunc}, where $\lambda_k$ are the characteristic roots of the characteristic equation associated to Eq.~\eqref{eq:diff_eq}, see Eq.~\eqref{eq:char_eq} below, and $d_k$ can be arbitrary coefficients due to linearity. This observation is crucial for inverting Eq.~\eqref{eq:kappa_n_trunc} and extracting the zeros from the cumulants as we explain below. We note that Eqs.~(\ref{eq:kappa_n_trunc},\ref{eq:diff_eq}) are exact only if the Loschmidt amplitude $\mathcal{Z}(\tau)$ is a polynomial with $m$ distinct zeros. In this case, the method provides exactly all the zeros of $\mathcal{Z}(\tau)$, independently of the cumulant orders used. In the general case, where $\mathcal{Z}(\tau)$ is an entire function, the method provides approximate zeros $\tau_k^{(n)}$ that depend on the cumulant orders and converge to the exact zeros for $n\to \infty$. Associated to the approximate zeros, the method provides also a sequence of approximate multiplicities $d_k^{(n)}$ for each $k$, which converges to the exact multiplicity $d_k$. 

The first step of the method is to compute the coefficients $a_{l=1,\dots,m}$ in Eq.~\eqref{eq:diff_eq}.
This can be done by solving a linear system of $m$ equations, which requires the knowledge of $2m$ consecutive cumulants ($\kappa_l$, with $n-m\leq l \leq n+m-1$) and takes the form
 \begin{widetext}
\begin{equation}
\label{eq:aj_sys}
\begin{pmatrix}
    \kappa_{n-1} & \kappa_{n-2} & \dots & \kappa_{n-m+1}  & \kappa_{n-m} \\
    \kappa_{n} & \kappa_{n-1} & \dots & \kappa_{n-m+2}  & \kappa_{n-m+1} \\
    \vdots & \vdots & \ddots & \vdots & \vdots \\
    \kappa_{n+m-3} & \kappa_{n+m-4} & \dots & \kappa_{n-1}  & \kappa_{n-2} \\
    \kappa_{n+m-2} & \kappa_{n+m-3} & \dots & \kappa_{n}  & \kappa_{n-1}
\end{pmatrix}
\begin{pmatrix}
a^{(n)}_1 \\ a^{(n)}_2 \\ \vdots \\a^{(n)}_{m-1} \\a^{(n)}_{m}
\end{pmatrix} =
\begin{pmatrix}
\kappa_n \\ \kappa_{n+1} \\ \vdots \\ \kappa_{n+m-2} \\ \kappa_{n+m-1} 
\end{pmatrix}.
\end{equation} 
The square matrix on the left hand side is a Toeplitz matrix. With the notation $a_l^{(n)}$, we emphasize that the coefficients of the linear difference equation obtained from Eq.~\eqref{eq:aj_sys} depend on the cumulant orders used, since Eqs.~(\ref{eq:kappa_n_trunc},\ref{eq:diff_eq}) are approximations in general.

The second step is to solve the characteristic equation associated to the linear difference equation~\eqref{eq:diff_eq},
\begin{equation}
\label{eq:char_eq}
\lambda^m - a^{(n)}_1\lambda^{m-1} -a^{(n)}_2\lambda^{m-2} -\ldots - a^{(n)}_{m-1}\lambda -a^{(n)}_m=0,
\end{equation}
which is a polynomial equation in $\lambda$. This provides the $m$ characteristic roots $\lambda^{(n)}_{k=0,\dots,m-1}$ of the difference equation~\eqref{eq:diff_eq}. Then, the approximate zeros are obtained from the relation $\tau^{(n)}_k = \tau +1/\lambda^{(n)}_k$. In Eq.~\eqref{eq:aj_sys}, we have suppressed the dependence of the cumulants $\kappa_n(\tau)$ on the basepoint $\tau$ at which they are calculated, see Eqs.~(\ref{eq:cumulant_series}, \ref{eq:kappa_n_trunc}). However, we emphasize that the approximate zeros $\tau^{(n)}_k$ depend in general not only on the cumulant orders used in Eq.~\eqref{eq:aj_sys}, but also on the basepoint $\tau$. As an example, by solving explicitly Eqs.~(\ref{eq:aj_sys},\ref{eq:char_eq}) in the case $m = 2$, one obtains 
\begin{equation}
\begin{split}
(\tau^{(n)}_0-\tau)+(\tau^{(n)}_1-\tau)&=\frac{\kappa_{n}\kappa_{n-1} -\kappa_{n-2} \kappa_{n+1}}{\kappa_{n}^2-\kappa_{n+1}\kappa_{n-1}},\\
(\tau_0^{(n)}-\tau)(\tau_1^{(n)}-\tau)&=\frac{\kappa_{n-1}^2-\kappa_{n}\kappa_{n-2}}{\kappa_{n}^2-\kappa_{n+1} \kappa_{n-1}},
\end{split}
\end{equation}
where $\tau_0^{(n)}$ and $\tau_1^{(n)}$ are two sequences that converge for $n\to \infty$ to the two closest zeros $\tau_0$ and $\tau_1$ ($\neq \tau_0^*$ in general). 

The third and final step is the calculation of the coefficients $d_k^{(n)}$ by solving the following linear system of equations,
\begin{equation}
\label{eq:dj_sys}
\begin{pmatrix}
    1         & 1         & \dots & 1   & 1 \\
    \lambda_0 & \lambda_1 & \dots & \lambda_{m-2} & \lambda_{m-1} \\
    \vdots & \vdots & \ddots & \vdots & \vdots \\
    \lambda_0^{m-2} & \lambda_1^{m-2} & \dots & \lambda_{m-2}^{m-2} & \lambda_{m-1}^{m-2} \\[0.3em]
    \lambda_0^{m-1} & \lambda_1^{m-1} & \dots & \lambda_{m-2}^{m-1} & \lambda_{m-1}^{m-1}
\end{pmatrix}
\begin{pmatrix}
d^{(n)}_0\lambda_0^n \\ d^{(n)}_1\lambda_1^n \\ \vdots \\d^{(n)}_{m-2}\lambda_{m-2}^n \\[0.3em]
 d^{(n)}_{m-1}\lambda_{m-1}^n
\end{pmatrix} =
\begin{pmatrix}
\kappa_n \\ \kappa_{n+1} \\ \vdots \\ \kappa_{n+m-2} \\[0.3em]
 \kappa_{n+m-1} 
\end{pmatrix}\,,
\end{equation}
\end{widetext}
where we have dropped the superscript from $\lambda_k^{(n)}$ in the above equation to ease the notation. The matrix on the left-hand side is a Vandermonde matrix, which is invertible, if all the $\lambda^{(n)}_k$ are distinct. Again, we emphasize that the approximate multiplicities $d_k^{(n)}$ depend on the chosen cumulant orders and on the basepoint, moreover, they are not exactly integers in general since Eqs.~(\ref{eq:kappa_n_trunc},\ref{eq:diff_eq}) are approximations. The remarkable property of the coefficients $d_k^{(n)}$ obtained from Eq.~\eqref{eq:dj_sys} is that they converge to the respective multiplicities $d_k^{(n)}\to d_k$ for $n\to \infty$ together with the approximate zeros $\tau_k^{(n)}\to \tau_k$. We use this fact to select the approximate zeros which are the best approximations of the exact zeros when the basepoint is varied as discussed in App.~\ref{app:errors}.

\section{Extracting Loschmidt zeros - numerical convergence and error estimates}
\label{app:errors}

In order to obtain accurate approximations of the exact zeros, one can increase the cumulant order (the parameter $n$ in Eq.~\eqref{eq:aj_sys}) to observe the convergence of the approximate zeros $\tau_k^{(n)}$. However, we use a different procedure in our work, which is almost automatic and has proven particularly effective for investigations of dynamical quantum phase transitions. The idea is based on the fact that the approximate multiplicities obtained from Eq.~\eqref{eq:dj_sys} converge to the exact multiplicities $d_k^{(n)} \to d_k$ concomitantly with the approximate zeros $\tau_n^{(n)}\to \tau_k$.
Indeed, in the case where the exact zeros $\tau_k$ and their multiplicities $d_k$ are known in advance (for instance the Kitaev chain with $V = 0$), we have observed that the distance $|\tau_k^{(n)}-\tau_k|$ is approximately linearly proportional to $|d_k^{(n)} -d_k|$. This is an extremely useful fact that allows us to automatically select the $\tau_k^{(n)}$ which are good approximations of the exact zeros. This is done by retaining only the pairs $(\tau_k^{(n)},\, d_k^{(n)})$ for which $|d_k^{(n)} - \ell|< r$, where $r$ is a fixed threshold and $\ell$ is a chosen integer. This condition appears to be necessary but not sufficient to guarantee that $\tau_k^{(n)}$ is a good approximation of $\tau_k$. Indeed, it can occur that the approximate zero of a pair $(\tau_k^{(n)}, d_k^{(n)})$ is not a good approximation of any exact zero even if the condition $|d^{(n)}_k-\ell|< r$ is satisfied for some integer $\ell$. However, this kind of false positives are in practice quite rare for small enough $r$ and can be easily detected and discarded by applying the method at different basepoints, as explained below.
The variant of the method in the case of a pair of conjugate zeros ($\tau_0 = \tau_1^*$, $m =2$) has been applied earlier to study critical phenomena in classical equilibrium problems \cite{Deger:2018,Deger:2019,Deger:2020,Deger:2020b}, and here we have extended the method to an arbitrary number of zeros, which are not necessarily pairwise conjugate. These are crucial advancements that have allowed us to apply the cumulant method to the study of dynamical quantum phase transitions. The use of the coefficients $d_k^{(n)}$ for selecting the best approximate zeros is also a new technique, which makes the method very practical and efficient for the prediction of dynamical quantum phase transitions and is used for the first time in this work.

The results shown in Figs.~\ref{fig:kitaev}-\ref{fig:spin1_2} have been obtained by evolving the initial state along the imaginary axis in the complex $\tau$-plane (path A in Fig.~\ref{fig:kitaev}\textbf{a}). In the case of the Kitaev chain, we perform the time evolution also along path B, see Fig.~\ref{fig:kitaev}\textbf{b}, a straight path starting at $\tau = 0$ and ending at $\tau = -4 + 6i$. In this way we can map out zeros in different regions of the complex plane. The intermediate values of $\tau$ along the evolution path are the basepoints at which the cumulants are computed. We use a fine grid in which adjacent basepoints are at a distance $\delta\tau = 0.001$. For each basepoint $\tau$ we apply Eqs.~\eqref{eq:aj_sys},~\eqref{eq:char_eq} and~\eqref{eq:dj_sys} using cumulants $\kappa_n(\tau)$ from $n =9$ to $n = 22$, and we obtain $m=7$ distinct approximate zeros $\tau_{k = 0,\dots,6}^{(j)}$ and corresponding coefficients $d_{k = 0.,\dots,6}^{(j)}$. Here we change the notation slightly: the superscript in the pair $(\tau_{k}^{(j)},d_{k}^{(j)})$ labels the distinct basepoints here. This is because the cumulant orders used in the method are kept fixed in this case and only the basepoint is varied along the path.

As explained above, we select only the pairs $(\tau_{k}^{(j)},d_{k}^{(j)})$ for which $|d_{k}^{(j)}-1| < r = 0.01$. We have not found zeros with multiplicity $\ell > 1$ in the systems considered in our work. After this selection step, one can visually verify that the approximate zeros tend to agglomerate in well separated clusters. In the case where the Loschmidt amplitude can be computed analytically and the exact zeros are known (the Kitaev chain with $V = 0$~\cite{Heyl:2013}, Fig.~\ref{fig:kitaev}\textbf{a}-\textbf{c}), one can also see that each cluster corresponds to one of the exact zeros. We use the standard $k$-means++ clustering algorithm~\cite{Arthur2007} to classify the approximate zeros in distinct clusters. This algorithm requires to introduce manually the number of clusters, which can be estimated by visual inspection. Each cluster obtained in this way typically consists of hundreds or thousands of pairs $(\tau_{k}^{(j)},d_{k}^{(j)})$. Clusters with less then ten pairs are discarded to eliminate any false positives.

Finally, within each cluster we select the pair $(\tau_{k}^{(j)},d_{k}^{(j)})$ for which $|d_k^{(j)}-1|$ takes its smallest value. The approximate zero $\tau_{k}^{(j)}$ of this pair gives the best estimate of the location of one zero. Indeed, we have observed in exactly solvable cases that the same pair minimizes also the distance $|\tau_k^{(j)} -\tau_k|$ from the closest exact zero. Notice that one can generally resolve more than $m$ zeros in a single run of time evolution by using the above procedure since different zeros are resolved for different basepoints.

The standard deviation $s$ of the real and imaginary parts of the approximate zeros within each cluster provides a rough estimate of the error, i.e. the distance from the exact zero. Typically, we obtain $s\approx 10^{-3}$ using the procedure presented above. By comparison with the exact solution we have verified that this is a reasonable estimate or even an overestimation in most cases. Another way to estimate the error is to compare the zeros obtained by evolving along different paths as in Fig.~\ref{fig:kitaev}\textbf{a}-\textbf{c}. Often the same exact zero can be resolved by using both paths and thus one obtains two different estimates of its location whose distance is an estimate of the error. The error estimate obtained in this way turns out to be essentially the same as the one obtained from the cluster standard deviation. An error of order $10^{-3}$ is not visible on the scale of Figs.~\ref{fig:kitaev}-\ref{fig:spin1_2}, since it is an order of magnitude smaller than the size of the markers. Therefore, the fact that in the interacting case the zeros seem not to organize in well defined lines (in contrast to the non-interacting case in Fig.~\ref{fig:kitaev}\textbf{a}-\textbf{c}) has to be entirely attributed to finite-size effects and not to the approximate nature of the zeros obtained with our method.

\section{Krylov subspace method}
\label{app:Krylov_method}

The evolution along a straight path in the complex $\tau$-plane is performed by using the standard Krylov subspace method~\cite{Park1986,Paeckel2019}. A time step $\delta\tau$ in the evolution is performed by first computing an orthonormal basis $B = \{ \ket{v_0} = \ket{\Psi(\tau)},\ket{v_1},\dots,\ket{v_{N_{\rm vec}}}\}$ of the Krylov subspace 
\begin{equation}
\label{eq:Krylov}
    \mathcal{K}_{N_{\rm vec}}=\mathrm{Span}\{ \mathcal{\hat H}  ^n\ket{\Psi(\tau)}|\,\, n = 0,\dots,N_{\rm vec} \}\,.
\end{equation}
We use the QR decomposition to compute the orthonormal basis of $\mathcal{K}_{N_{\rm vec}}$ to ensure that there is no loss of orthogonality as in the standard Lanczos algorithm~\cite{Paeckel2019}. Then, the approximate time-evolved state is obtained as 
\begin{equation}
\label{eq:exp_approx}
\ket{\Psi(\tau+\delta\tau)} \simeq e^{-\delta\tau\hat{H}_{\rm eff}}\ket{\Psi(\tau)}\,,
\end{equation}
where $\hat{H}_{\rm eff}$ is the effective Hamiltonian, that is an operator which acts on the Krylov subspace and whose matrix elements are given by $\bra{v_i}\mathcal{\hat H}\ket{v_j}$ with $\ket{v_i} \in B$. It is represented by a square matrix of dimension $N_{\rm vec}+1$ whose exponential is easy to evaluate since we take $N_{\rm vec} = 8$ in our case. As suggested in Section~5 of Ref.~\cite{Paeckel2019}, the effective Hamiltonian is forcibly set to be a tridiagonal matrix to improve stability.

In order to estimate the accuracy of Eq.~(\ref{eq:exp_approx}) we perform the time evolution on two Kyrlov subspaces $\mathcal{K}_{N'_{\rm vec}}$ and $\mathcal{K}_{N''_{\rm vec}}$ with $N'_{\rm vec}+1 = N''_{\rm vec}\leq N_{\rm vec}$ and compute the distance between the approximate evolved states $d = \|\ket{\Psi'(\tau+\delta\tau)} - \ket{\Psi''(\tau+\delta\tau)} \|_2$. If $d < 10^{-10}$ the state 
$\ket{\Psi''(\tau+\delta\tau)}$ is stored and used to evaluate the moments of the Loschmidt amplitude according to $\langle \mathcal{\hat H}^n \rangle_{\tau + \delta\tau} = \bra{\Psi_0} \mathcal{\hat H}^n \ket{\Psi''(\tau +\delta\tau)}$. On the other hand, if for a given $\delta \tau$ the condition is not satisfied even for $N''_{\rm vec} = N_{\rm vec}$, the Krylov subspace $\mathcal{K}_{N_{\rm vec}}$ in Eq.~\eqref{eq:Krylov} is recomputed using as the seed state the last stored vector $\ket{\Psi''(\tau+\delta\tau)}$. The whole process is repeated up to the desired final $\tau$.

An important advantage of the Krylov time evolution algorithm described above is that intermediate values of $\tau$ along the evolution path are easily accessible at a negligible computational cost. 
We take advantage of this fact to compute the cumulants on a fine grid of basepoints along the evolution path.
The computationally expensive part of the method is the calculation of the Krylov subspace~\eqref{eq:Krylov}. In order to compute the moments and the cumulants, a larger Krylov subspace ($N_{\rm vec} = 22$ in our case) has to be calculated only at $\tau = 0$. This is a small numerical overhead since the Krylov subspace has to be recalculated many times during the evolution. With the specific parameters given above, we find by comparison with exactly solvable cases that the cumulants are computed with a relative precision of $10^{-9}$, which is sufficient for our purposes. 

\section{Twisted boundary conditions}
\label{app:boundary_conditions}

The twisted boundary conditions that we employ are known as a tool to filter out finite-size effects. For example, twisted boundary conditions have been employed to analyze energy-level crossings related to quantum phase transitions in spin-1 systems \cite{Fath:1993}. Here we discuss how these boundary conditions can also help to gauge finite-size effects in the Loschmidt zeros. For the Haldane chain, Eq.~\eqref{eq:spinone}, the spectrum exhibits a $1/L$ dependence on the boundary conditions \cite{Fath:1993}. Thus, the Loschmidt zeros, determined by the spectrum of the Hamiltonian, are independent of the boundary conditions for sufficiently large system sizes. This is a reasonable assumption for most naturally occurring systems. However, one may in principle envision situations that deviate from this generic behavior, and one may even construct specific examples that do \cite{torre2021odd}. On the other hand, in finite systems, the many-body energies $E_i(\Phi)$ depend on the twist angle $\Phi$, in particular the highly excited states. The Loschmidt zeros inherit this property -- a single zero, say $\tau_k$, gives rise to a multiplet of zeros $\tau_k(\Phi)$ corresponding to different values of $\Phi$ as seen in Fig.~\ref{fig:spin1_1} (in particular, see panel \textbf{a1}) and in Fig.~\ref{fig:spin1_2}. Fortunately, we observe a regular pattern, which can be employed as a diagnostic tool. Specifically, all $E_i(\Phi)$ are even functions of $\Phi$, having their extreme values at $\Phi=0$ or $\Phi=\pi$. As seen in Fig.~\ref{fig:spin1_1}\textbf{a1}, the boundary points of the multiplets of zeros exactly correspond to $\Phi=0$ or $\Phi=\pi$. In the thermodynamic limit, these multiplets converge to a single point, $\tau_k$, for all angles $\Phi$. 

A natural question is how one can obtain the best approximation for the thermodynamic line of zeros. It is reasonable to expect that the line is situated between the extreme zeros at the twist angles $\Phi=0$ and $\pi$. The curve of zeros corresponding to a general value of $\Phi$ exhibits size-dependent oscillations around the mean value as seen in Fig.~\ref{fig:spin1_1}\textbf{a1} by the zig-zag line corresponding to $\Phi=0$. However, these finite-size oscillations are suppressed for zeros corresponding to the twist angle $\Phi=\pi/2$, which form a smooth curve. This observation suggests that the twist angle $\Phi=\pi/2$ is special, and that it is the best approximation for the thermodynamic line of zeros. This prescription is independently supported by the remarkable agreement between the critical times that we obtain and the results of Ref.~\cite{Hagymasi:2019}, which are also indicated in Fig.~\ref{fig:spin1_1}. A similar pattern is observed for the quench from the Haldane to the N\'eel phase in Fig.~\ref{fig:spin1_2}, providing further support for approximating the thermodynamic lines of zeros using the zeros corresponding to the boundary condition $\Phi=\pi/2$.

\bibliography{biblio}

\end{document}